# Brilliant source of 19.2 attosecond soft X-ray pulses below the atomic unit of time.

Fernando Ardana-Lamas[1,†], Seth L. Cousin[1,†], Juliette Lignieres[1], Jens Biegert[1,2,*]

[1] ICFO - Institut de Ciencies Fotoniques, The Barcelona Institute of Science and Technology, 08860 Castelldefels (Barcelona), Spain.
[2] ICREA, Pg. Lluís Companys 23, 08010 Barcelona, Spain.
[*] Correspondence to: jens.biegert@icfo.eu
[†] These authors contributed equally

**Abstract**
Electronic correlations occur on attosecond timescales, dictating how chemical bonds form, energy flows, and materials respond to light. Capturing such many-body processes requires light pulses of similar duration. The soft X-ray water window is vital because it encompasses the principle absorption edges of carbon, nitrogen, and oxygen that underpin chemistry, biology, and materials science. However, generating and characterising isolated attosecond pulses that reach into the soft X-ray water window has remained a challenge.
We addressed this need and report an isolated attosecond soft X-ray pulse with a duration of 19.2 attoseconds. This pulse reaches into the water window and is shorter than the atomic unit of time (24.2 as). The pulse is supported by a spectrum centred at 243 eV, extending up to 390 eV, and crossing the carbon K-edge with record photon flux: $4.8 \times 10^{10}$ photons per second overall and $4.1 \times 10^{9}$ photons per second in a 10% bandwidth at the carbon K-shell edge (284 eV). Such an extremely short soft X-ray pulse combines extreme temporal resolution with a coherent ultrabroadband soft X-ray spectrum, opening new opportunities to study electron dynamics in atoms, molecules, and solids, disentangle many-body interactions in correlated systems, and follow non-adiabatic energy flow in molecular complexes. Our results establish a new benchmark for table-top attosecond technology and lay the foundation for its widespread application in science and technology.

**INTRODUCTION**

Excitation, scattering, and electron relaxation are crucial processes that control how matter interacts with light. Their timing influences how chemical bonds form or break, how charge and energy move, and how the functions in molecules and materials develop. Understanding these dynamics requires attosecond resolution, as electronic excitations and dynamics occur on timescales of tens of attoseconds (*1*). Since the first real-time observation of the femtosecond lifetime of an M-shell vacancy in krypton in 2002 (*2*), attosecond science has enabled groundbreaking studies of electron tunnelling (*3*), photoemission (*4, 5*), ultrafast charge transfer in molecules (*6–11*), electron localisation during dissociation (*12–14*), electronic dynamics in condensed matter (*15–19*) and probing many-body dynamics in condensed matter (*20, 21*).

Extending attosecond pulses into the soft X-ray (SXR, > 124 eV) regime is particularly powerful because core-level spectroscopy offers element specificity and site selectivity. This capability allows attosecond-resolution investigations of intra-atomic and intra-molecular energy transfer (*22–24*), charge-induced structural rearrangements (*25, 26*), transient variations in chemical reactivity (*27*), non-adiabatic dynamics (*28–30*), photodamage of organic matter (*31*), and carrier and exciton dynamics (*16, 20, 32*). Significantly, SXR attosecond pulses permit access to core-shell absorption edges vital to chemistry and biology, such as the carbon, nitrogen, and oxygen K and L-edges, thus providing a novel means to explore the complex many-body interactions between photons, charge carriers, and nuclei (*20*, *21*, *33*).

Over the past decade, short- and mid-wave infrared drivers have advanced high-harmonic generation (HHG) (*34*, *35*) into the SXR domain (*36–41*). Early HHG research using Ti:Sapphire lasers led to the production of keV photons (*42*), while the ponderomotive scaling of HHG (*43*) with longer wavelength drivers enabled generation within the water window (284-530 eV; (*39*)) and up to the keV range (*36*). However, it required substantial progress (*39*, *44*) to overcome the unfavourable wavelength scaling of the single-atom response ($\lambda^{-5}$ to $\lambda^{-9}$) (*45–47*) by understanding ionisation dynamics at unprecedented gas pressures (*38*, *44*, *48*) to achieve efficient phase-matched HHG (*49–52*). Such conditions require pressures of several to tens of bars, and can be obtained using capillaries or carefully designed target systems that support stable, effusive gas targets. These innovations have enabled the long-sought new opportunities, such as the first coherent SXR spectroscopy (*17*, *37*, *53*). Simultaneously, HHG in the SXR range has led to expectations of generating even shorter, isolated attosecond pulses (*54–59*), which face new challenges due to a significantly reduced photon flux in the SXR (*47*), as well as measurement difficulties (*60*, *61*). Several studies have reported (*60*, *62–64*) achieving the SXR regime with pulse durations close to the atomic unit of time (24 as). Nonetheless, key challenges in such streaking measurements (*65*, *66*) of an SXR attosecond pulse include emission from multiple shells of the measurement gas (*60*) and limitations of retrieval methods (*67*, *68*) to extract the electric field envelope of the SXR pulse accurately. However, it is well known that a combination of time-stationary and non-stationary filters must be used to retrieve the electric field amplitude of the attosecond pulse; otherwise, only the spectrum or an undersampled temporal trace can be obtained.

Chronocyclic techniques, such as the FROG-CRAB (*67*) method, are well-established methods that combine time-stationary and time-non-stationary filters to extract an electric field envelope (*69*). FROG-CRAB, however, relies in its original form on the central momentum approximation (CMA); therefore, it must be used cautiously when dealing with broad SXR bandwidths. Frequency-domain interferometric methods, like RABBITT (*70*), PROOF (*71*, *72*), and iPROOF (*73*), avoid the CMA but are limited to weak streaking fields and long infrared pulses. As a result, achieving an accurate reconstruction of extremely short, i.e., broadband, SXR pulses alongside their streaking fields has remained a challenge. One solution to this problem is the Volkov-transform generalised projection algorithm (VTGPA) (*74*), which formulates the retrieval as a Fourier-transform-free least-squares optimisation and bypasses the CMA. The VTGPA further incorporates dipole transition matrix elements. This method reduces computational demands, improves robustness to noise, and converges in about 20 minutes on modern computing hardware, making it a practical tool for characterising SXR attosecond experiments.

Using this method, we measured an isolated attosecond pulse with a duration below the atomic unit of time (24.2 as), reaching 19.2 attoseconds. The central photon energy and spectral bandwidth surpass previous demonstrations, setting a new benchmark for attosecond pulse generation and characterisation.

The availability of such a pulse marks a significant advance for attosecond science and ultrafast investigations due to its short duration and fully coherent bandwidth for spectroscopy (*33*). Covering the key absorption edges of light elements, they enable element-specific attosecond spectroscopy of inner-shell relaxation, Auger decay, and correlated carrier–nuclear dynamics in gases, liquids, and solids (*33*). They also open a new chapter for chemical research, from ultrafast charge migration in biomolecules to structural transitions in condensed matter and light-induced phase changes in correlated materials. By pushing attosecond science below the atomic time unit and into the water window, this work provides a vital tool to study matter on its most fundamental temporal and spatial scales.

## RESULTS AND DISCUSSION

### Attosecond SXR Measurement Setup

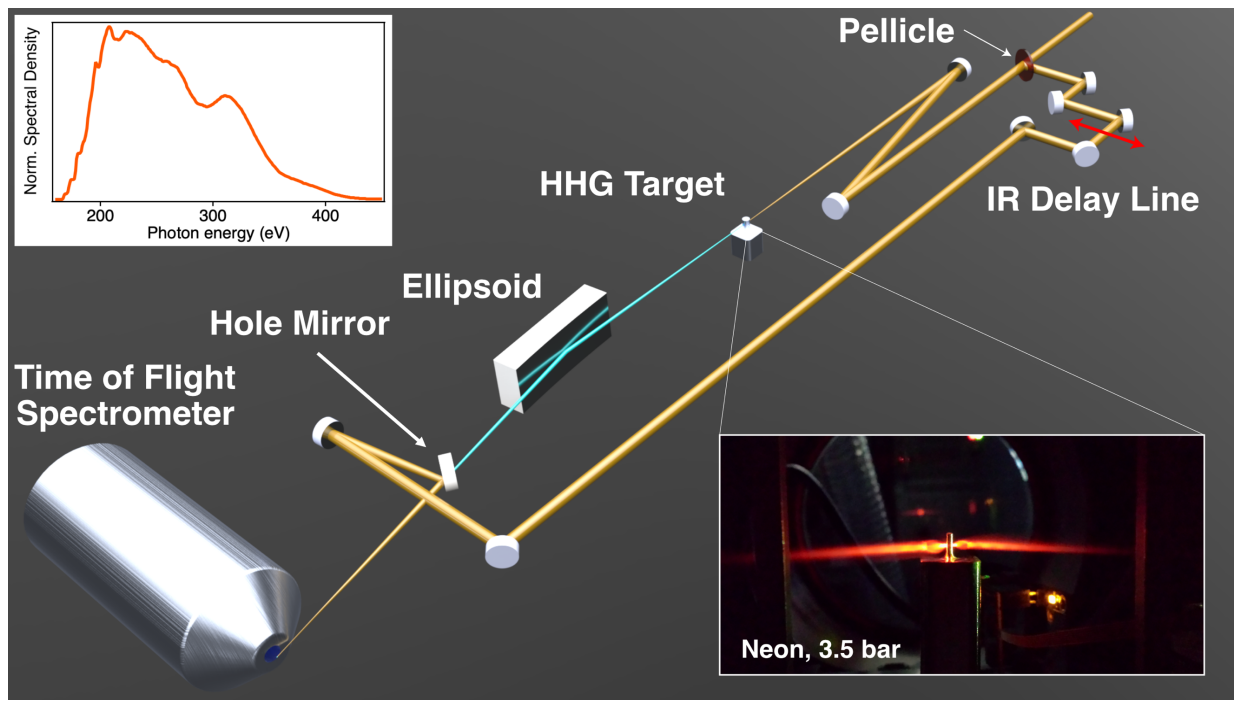

Figure 1: **Attosecond pulse generation and measurement.** Schematic outline of the measurement setup. The 12-fs CEP-stable pulse at 1850 nm is split by a pellicle beam, with a major part of the energy used to generate the SXR attosecond pulse in an effusive "T" shaped target described in Ref. (*37*). The inset (bottom right) shows a photograph of the target when operating with 3.5 bar neon. The emerging attosecond pulse is focused into a TOF spectrometer, and the remainder of the short pulse is used as a streaking field.

Here we briefly describe the experimental setup. The driving source was a home-built, cryogenically cooled, two-stage Ti:Sapphire amplifier that delivered stable 40-fs, 7-mJ pulses at 1 kHz. These pulses seeded a TOPAS-HE (LightConversion) optical parametric amplifier (OPA), where three stages of white-light-seeded amplification generated passively CEP-stabilised idler pulses of 45 fs duration centred at 1850 nm. The pulses were spectrally broadened in a gas-filled hollow-core fibre and subsequently compressed in bulk, resulting in 0.4-mJ, 12-fs, CEP-stable pulses. A slow feedback loop mitigated long-term CEP drifts, ensuring stability over extended timescales. A measurement over 1 hour yielded an rms stability of 88.8 mrad. These pulses drove the attosecond streaking beamline, where most of the energy was used to generate the attosecond SXR pulse, while a smaller portion was used as the streaking field; see Fig. 1 for an illustration.

The high-energy pulse was focused to a spot size of 54 μm and reached an intensity of $4.3 \times 10^{14}$ W/cm² in a free-space neon target with a backing pressure of 3.5 bar and an interaction region shorter than 1 mm. Under these conditions, soft X-rays were generated well into the water window, delivering a total flux of $4.8 \times 10^{10}$ photons/s across the spectral range from 200 eV to 380 eV. The top left inset in Fig. 1 shows a typical SXR spectrum, and the bottom right inset shows a photograph of the high-pressure target.

The generated soft X-rays were refocused into a time-of-flight (TOF) spectrometer using a grazing-incidence ellipsoidal mirror (Zeiss) with ion-beam-polished surfaces of less than 0.5 nm roughness across the 260 × 50 mm² optic. A gold mirror with a 3-mm through-hole at 45° enabled co-alignment of the 1850-nm streaking pulses (reflected) with the X-rays (transmitted). Beam steering was performed with motorised mirrors under vacuum, and an actuator-controlled sampling mirror allowed real-time beam profiling. Collinearity was verified when the beams from both arms overlapped on the hole and remained aligned in the far field.

The interferometric stability of the streaking setup was qualitatively confirmed by observing the interference fringes of the 1850-nm beams at the TOF gas jet. Excellent fringe stability was achieved without active stabilisation, thanks to isolating the vacuum chambers from the optical breadboards. No infrared attenuation filter was employed, maximising the soft X-ray flux at the target. Control experiments verified that residual infrared did not affect the streaking trace; see Ref. (*60*): its intensity was three orders of magnitude below the ionisation threshold, and comparisons of photoelectron spectra recorded with and without infrared-blocking filters (200 nm carbon + 200 nm chromium) revealed no energy shift, only a reduction in the signal.

**Attosecond SXR Pulse Measurement**

Now we turn to the streaking measurement. We previously described the characterisation of attosecond SXR pulses generated in neon at 3.5 bars using streaking measurements in krypton (*60*). In that work, the FROG-CRAB retrieval method was employed; however, due to its limitations for the broad SXR bandwidth produced, we restricted ourselves to only reporting a conservative upper bound of 322 as for the pulse duration. Notably, the retrieval already indicated the presence of isolated attosecond pulses with durations of 23.1 as, hinting at the ultrashort regime we now confirm.

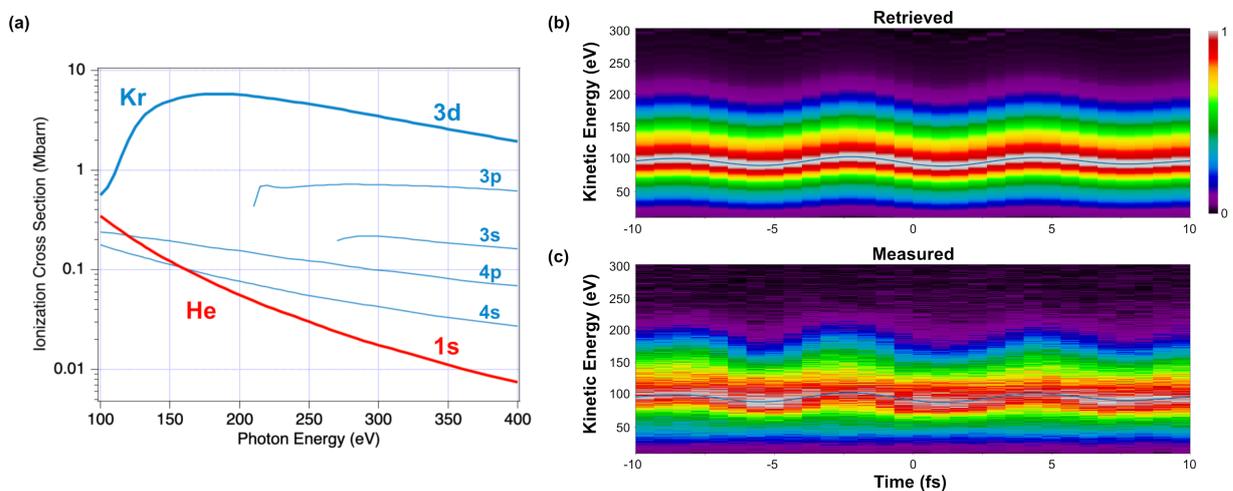

Figure 2: **Attosecond streaking in Krypton.** (a) Photoionisation cross sections for helium and krypton. Helium would provide direct mapping, but it is prohibitive due to the extremely low cross-section. We used krypton with a binding energy of $E_B = 94$ eV and ensured that the additional shells do not pollute the measured trace shown to the right. (b) and (c) Retrieved and measured streaking traces in krypton. The black line represents the vector potential.

We briefly outline the measurement conditions. A portion of the IR pulse energy was split off to act as the streaking field, reaching a peak intensity of $3.2 \times 10^{11}$ W/cm² at focus. The TOF spectrometer (Stefan Kaesdorf) was operated in either electron or ion collection mode and included a 461-mm drift tube, providing a resolution of T/ΔT = 100, corresponding to E/ΔE ≈ 50. An Einzel lens focused charged particles onto a microchannel plate without changing their energies; optimal focusing was achieved when the lens voltage was approximately five times the expected electron energy. For a central energy of 200 eV, the acceptance bandwidth was around 150 eV (FWHM). In this setup, the standard 30° full-cone acceptance was replaced by a spherical acceptance volume approximately 200 μm in diameter, requiring precise co-alignment of the soft X-rays, streaking infrared, and gas jet. Helium is ideal as the target gas since emission from only the 1s shell precisely defines the photon energy of emitted photoelectrons $E_{kin} = h\nu - E_B$; $h\nu$ is the SXR photon energy, and $E_B$ is the electron's binding energy relative to the vacuum level. However, the ionisation cross section of helium is more than two orders of magnitude lower than that of krypton, making krypton a practical choice for measuring SXR pulses — that is, for photon energies above 124 eV; see Fig. 2. We emphasise that careful assessment of the spectral distribution of photon flux across the attosecond spectrum is necessary. For example, if the photon flux mainly resides at 100 eV with only a tail crossing into the water window, measuring with helium results in a significantly skewed mapping of the attosecond pulse spectrum into the photoelectron spectrum.

We previously (*60*) performed a detailed analysis of the SXR-driven photoemission of krypton, which clearly ruled out significant contributions other than the Kr(3d) shell. The photoionisation cross section of the Kr(3d) shell at the centre of mass of the attosecond pulse spectrum at 243 eV is nearly ten times larger than that of the Kr(3p) shell and almost one hundred times larger than that of the 4p shell. At this energy, the asymmetry parameters of the Kr(3d) and Kr(3p) shells converge near unity and vary minimally across the pulse bandwidth. Thus, photoemission can be attributed predominantly to the Kr(3d) shell, with angular asymmetries posing no limitation. However, Auger decay is another source of unwanted electrons, with secondary contributions. We found that ionisation from the Kr(3d-1) shell produces Auger electrons with energies below 60 eV (*75*), well outside the energy window contributing to streaking, and with negligible intensity, at most 6% of the Kr(3d) signal. A second source could be Auger decay from the Kr(3p) shell. Analysis shows that the three relevant groups ($M_{2,3}$-$M_{4,5}N_{2,3}$, $M_2$-$M_{4,5}N_{2,3}$ and $M_3$-$M_{4,5}N_{2,3}$) amount maximally to 6% of the Kr(3d) signal. Lastly, the $M_{2,3}$–NN Auger lines account for only 0.66% of the Kr(3d) signal. These contributions are therefore negligible, confirming that the detected photoelectrons arise from single-photon ionisation of the Kr(3d) shell, providing a direct one-to-one mapping between the attosecond SXR pulse and the photoelectron TOF spectrum.

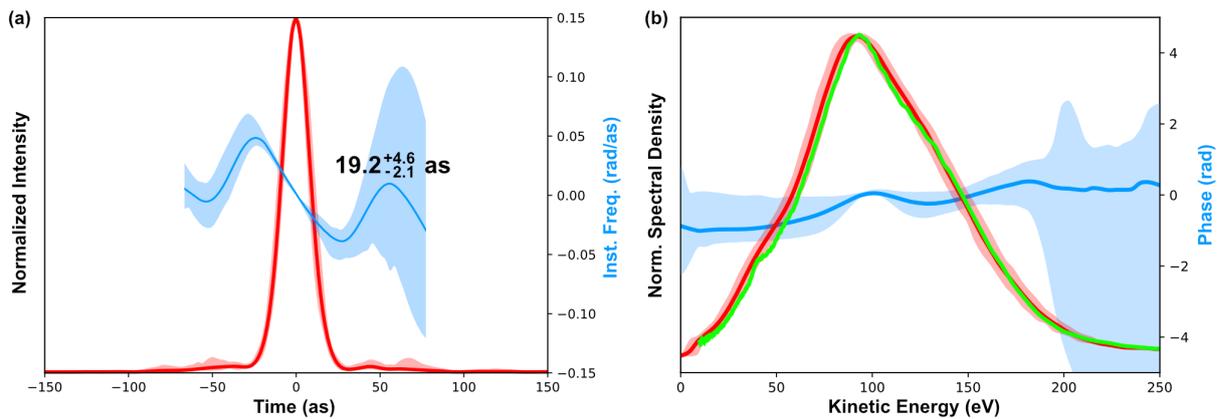

Figure 3: **Reconstructed attosecond SXR pulse.** (a) The intensity profile of the soft X-ray attosecond pulse had an average FWHM pulse duration of 19.2 (-2.1 / +4.6) as. The corresponding instantaneous frequency is also shown, revealing that the SXR pulse is down chirped with a slight residual phase. The shaded areas indicate the variations, and the thick lines the mean values. (b) The corresponding measured (green line) and retrieved kinetic energy spectra are shown in the right panel with the retrieved spectral phase.

Figure 2 shows the recorded and retrieved streaking traces. We used the Volkov-transform generalised projection algorithm (VTGPA) to retrieve the attosecond pulse. By framing the reconstruction as a least-squares projection in the Volkov-transformed domain, VTGPA (*74*) avoids the central momentum approximation and captures the complete physics of strong-field photoionisation, including dipole transition matrix elements. This allows for reliable retrieval of both the attosecond pulse and the streaking field, even for broadband soft X-ray continua and nonperturbative IR streaking fields. It is particularly effective for characterising our ultra-broadband attosecond SXR pulses. The VTGPA was repeated 30 times and yielded the average pulse duration of 19.2 (-2.1 / +4.6) attoseconds, well below the atomic unit of time of 24.2 as. Figure 3(a) shows the mean intensity profile and instantaneous frequency. Shaded areas indicate the spread of values. Figure. 3(b) shows the measured (green) and retrieved kinetic energy spectra with the retrieved spectral phase and spread of values indicated in shaded areas.

The retrieval started with a random guess and converged with an error of $\varepsilon = 100 \sum_t \sum_W |S_{ret} - S_{exp}|^2 \Delta W / N_t \sum_W \Delta W = 0.15\%$; where W is the kinetic energy, S is the spectra and energy steps. We further analysed the entire streaking trace by retrieving the attosecond pulse at each half-cycle interval of the streaking field. We observe consistent retrieval of similar pulse durations.

Regarding the residual uncompensated chirp, Fig. 4 illustrates the markedly different challenges for chirp compensation in the SXR regime compared to the XUV regime. The figure shows calculated transmission and GDD for a range of noble gases at 100 mbar pressure and typical foils with 100 nm thickness. We calculated these values based on tabulated data (*76*) with adjustment for temperature and pressure (*77*). The differences between the XUV (up to 124 eV) and SXR (above 124 eV) regimes are striking and illustrate the challenges in achieving appreciable chirp compensation without strongly modulating the coherent attosecond spectrum or reducing photon flux. Although straightforward post-generation chirp compensation by inserting materials or gases seems impractical in the SXR, earlier studies have shown that high-order harmonic generation (HHG) at long driving wavelengths, high intensities, and elevated gas pressures can demonstrate near-instantaneous temporal gating. For example, when the gas pressure was increased, Ref. (*78*) reported transitioning from attosecond pulse trains to isolated attosecond pulses—even with multicycle driving pulses. Similarly, our work (*38*, *44*) revealed that under multi-atmosphere HHG conditions, phase matching is inherently confined to a temporal window much shorter than half a cycle of the 1850 nm driving field. We note that some control may also be possible with an additional control field. A simple calculation shows that the intrinsic attochirp varies from 1654 $as^2$ to 6172 $as^2$ when decreasing peak intensity from $6.8 \times 10^{14}$ to $1.7 \times 10^{14}$ W/cm². The rapidly varying electric field envelope of the generating few-cycle pulse ramps through a much larger range of intensities and may provide additional chirp compensation.

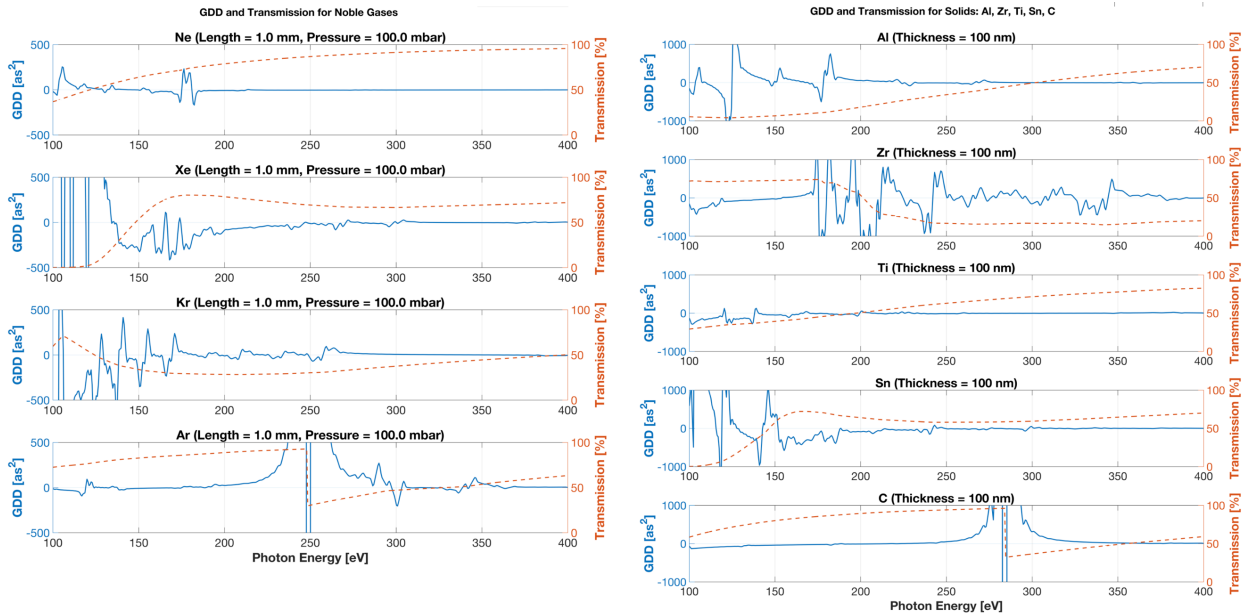

Figure 4: **Dispersion and transmission of XUV and SXR radiation.** We calculate the group delay dispersion (GDD) and the transmission based on reference data (*73*), which was adjusted for pressure and temperature (*74*). The left column gives values for noble gases, showing well-known resonances in the XUV spectrum for Kr and Xe. Data for a range of solid foils are shown in the right column. Note that a resonance causes a strong amplitude modulation of the attosecond pulse spectrum, but more importantly, it causes a phase modulation of the spectral phase.

Another critical aspect of HHG at multi-atmosphere pressure is understanding the gas dynamics, as it influences phase matching and gives valuable insight into target design and optimisation of photon yield. Figure 5 shows the results of our investigation, in which computational fluid dynamics simulations on ANSYS Fluent CFD with the K-epsilon turbulence model and a finite element size of 10 microns were matched with the gas jet shown in Fig. 5(top, left). The expansion of 3.5 bars of neon through a 200 micron orifice produces turbulent flow and shock fronts nicely reproduced by the simulations. The panels below (Fig. 5(a)-(d)) show the computed distribution of velocity reaching speeds up to 720 m/s, as well as the density and temperature distributions; also shown in Fig. 5(e,f). Based on these values, we calculated the chirp as a function of photon energy and found a GDD of -101.3 $as^2$ at 243 eV; see Fig. 5(g).

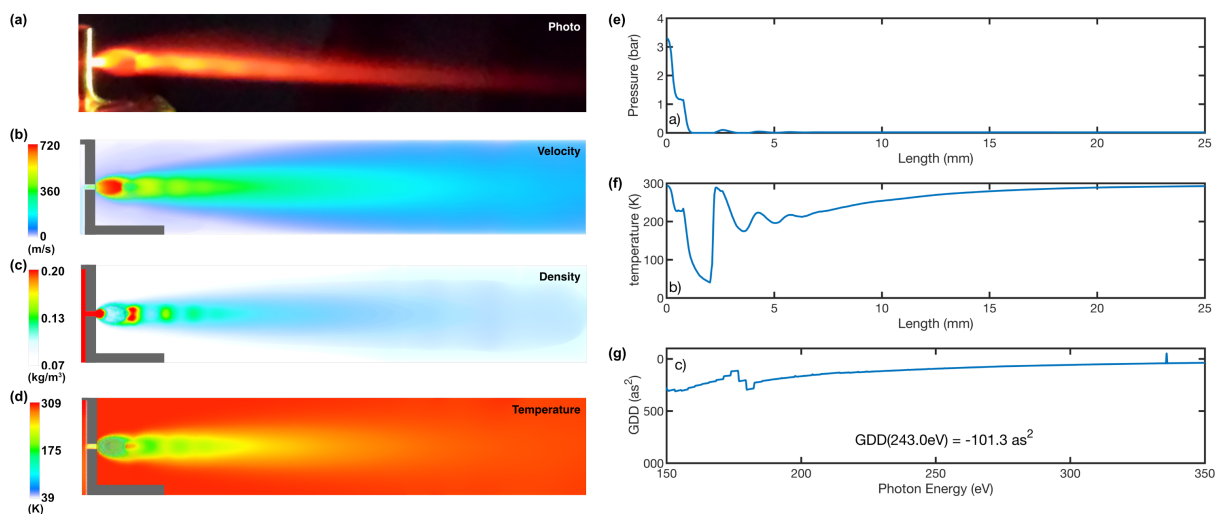

Figure 5: **Multi-bar gas dynamics and chirp in HHG.** (a) photograph of the gas jet. CFD simulations provided velocity (b), density (c) and temperature (d) of the HHG target. Lineouts of the CFD data along the beam propagation yield the pressure of neon (e), the temperature (f) and the GDD (g). The total transmission was 3%.

## CONCLUSION

We demonstrated the generation of coherent attosecond SXR pulses with a duration of 19.2 attoseconds, significantly shorter than the atomic unit of time. These pulses are centred at 243 eV with a spectrum extending to 390 eV and record photon flux of $4.8\times10^{10}$ photons per second. The photon flux in 10% bandwidth at the carbon K-shell edge at 284 eV is $4.1\times10^{9}$ photons per second. The sub-atomic time duration and the coherence of the ultrabroad spectrum offer exciting new opportunities to investigate atomic, molecular, and solid-state physics, such as inner and valence electron dynamics, or to disentangle many-body physics in correlated systems and non-adiabatic energy transfer in molecular complexes. Our results demonstrate the remarkable capabilities of table-top attosecond technology and lay the foundation for its widespread use in fundamental and applied science.

## ACKNOWLEDGMENTS

We gratefully acknowledge the contributions of all the authors of our original publication (*60*) on which this new work is based.

**Author contributions:** F.A.-L. and J.L. conducted the retrieval simulations, which F.A.-L. supervised; S.L.C. conducted fluid dynamic simulations and dispersion calculations; J.B. supervised the investigation and wrote the manuscript with contributions from all authors.

**Funding:** J.B. and group acknowledges financial support from the European Research Council for ERC Advanced Grant "TRANSFORMER" (788218), ERC Proof of Concept Grant "miniX" (840010), FET-OPEN "PETACom" (829153), FET-OPEN "OPTOlogic" (899794), FET-OPEN "TwistedNano" (101046424), MINECO for Plan Nacional PID2024-162757NB-I00; QU-ATTO, 101168628; AGAUR for 2021 SGR 01449, MINECO for "Severo Ochoa" (CEX2019-000910-S), Fundació Cellex Barcelona, the CERCA Programme/Generalitat de Catalunya, and the Alexander von Humboldt Foundation for the Friedrich Wilhelm Bessel Prize. JB also acknowledges Lasers4EU, which is funded by the European Union funds under HEU-GA 101131771.